\documentclass[12pt,fleqn]{article}
\usepackage{color,amsmath,epsfig}

\newcommand{\be}{\begin{eqnarray}}
\newcommand{\ee}{\end{eqnarray}}
\newcommand{\nee}{\nonumber\end{eqnarray}}
\newcommand{\nn}{\nonumber}
\newcommand{\noi}{\noindent}

\newcommand{\smaf}[2]  {{\textstyle \frac{#1}{#2} }}

\newcommand{\fig}[1] {\mbox{Fig.~\ref{fig:#1}}}

\def\b               {\beta}
\def\d               {\delta}

\def\G               {\Gamma}

\def\x               {\chi}

\def\ti              {\tilde}

\def\snu              {\ti \nu}
\def\stau              {\ti \tau}

\def\st              {\ti t}
\def\sb              {\ti b}

\def\chp             {\ti \x^+}
\def\chm             {\ti \x^-}
\def\nt              {\ti \x^0}
\def\sg              {\ti g}

\newcommand{\mst}[1]   {m_{\st_{#1}}}
\newcommand{\msb}[1]   {m_{\sb_{#1}}}

\newcommand{\mhp}      {m_{H^+}}

\begin{document}


\begin{center}

{\large\bf\boldmath
    CP asymmetries in charged Higgs boson decays in MSSM }\\[5mm]

E. Ginina\\[5mm]

{\it Contribution to the 
4th workshop "Gravity, Astrophysics, and Strings at the Black Sea", Primorsko, June 10-16, 2007}

\end{center}

\begin{abstract}
In the Standard Model with Minimal Supersymmetry, the Lagrangian contains complex parameters which lead to additional CP violation. We study CP violating asymmetries in the decays of the MSSM charged Higgs boson $H^\pm$, induced by loop corrections with intermediate SUSY particles, and perform analytical and numerical analysis. The decay rate asymmetry can go up to $25\%$ and the forward-backward asymmetry can reach up to $10\%$.
\end{abstract}


The CP--symmetry is broken in nature and this feature makes it an important tool for testing fundamental theories. An understanding of its non--conservation in particle physics may provide essential information about the mechanism, causing the dominance of the matter over antimatter in the observable universe. Generally, effects of CP--violation originate in non vanishing complex phases in the Lagrangian of the theory. In particular,
in the Minimal Supersymmetric Standard Model (MSSM), the Higgs mixing
parameter $\mu$ in the superpotential, the U(1) and SU(2) gaugino
mass parameters $M_1$ and $M_2$, and the trilinear couplings $A_f$
(corresponding to a fermion $f$) can have physical phases which cannot be rotated away
without introducing phases in other couplings~\cite{Dugan:1984qf}. The experimental upper bounds
on the electron and neutron electric dipole moments (EDMs)
constrain the phase of $\mu$, $\phi_\mu < {\cal
O}(10^{-2})$~\cite{Nath:dn} for a typical SUSY mass scale of the
order of a few hundred GeV, however the phases of the other parameters
mentioned above are practically unconstrained. The CP-violating
effects that might arise from the trilinear couplings of the first
generation $A_{u,d}$ are relatively small, as they are proportional
to $m_{u,d}$. However, the trilinear couplings of the third
generation $A_{t,b,\tau}$ can lead to significant effects of CP-violation, especially in top quark physics~\cite{Atwood:2000tu}.

In the following, we study CP violation in the decays of the
charged Higgs bosons $H^\pm$ within the MSSM. At tree level, there are three possible decay modes
of $H^+$  into ordinary particles: $H^+ \to t\bar b$, $H^+ \to
\nu\tau^+$ and $H^+ \to W^+h^0$, where $h^0$ is the lightest
neutral Higgs boson. Loop corrections due to a Lagrangian with complex
parameters lead to non zero decay rate asymmetry between the partial
decay widths of $H^+$ and $H^-$,
\begin{equation}
  \d^{CP} = \frac{\G\,(H^+\to ...)-\G\,(H^-\to ...)}
                 {\G\,(H^+\to ...)+\G\,(H^-\to ...)}\,,
\label{eq:defDCP}
\end{equation}
and that would be a clear signal
of CP violation. We consider such decay rate
asymmetries in MSSM with complex parameters for the quark decay
mode $H^+ \to t\bar b$ \, -- the asymmetry $\delta_{tb}^{CP}$
~\cite{we2}, and for the bosonic decay mode $H^+ \to W^+h^0$
~\cite{our} -- the asymmetry $\delta_{Wh^0}^{CP}$. For the quark
decay mode $H^+ \to t\bar b$ we go a step further by
including the decay products of the top quark. This allows to examine CP-violating asymmetries due to the polarization of the top quark ~\cite{second}. 
The top-quark decays before forming a bound state due to its large
mass, so that the polarization can be measured by the angular
distributions of its decay products. Moreover, the polarization is very
sensitive to CP violation. The considered CP violating
forward-backward (FB) and energy asymmetries are constructed by using angular or
energy distributions of the decay particles, following the formalism of \cite{katia}. This talk is entirely based on the results of \cite{our} and \cite{second}.

\begin{figure}[h!]
 \begin{center}
 \mbox{\resizebox{!}{4.3cm}{\includegraphics{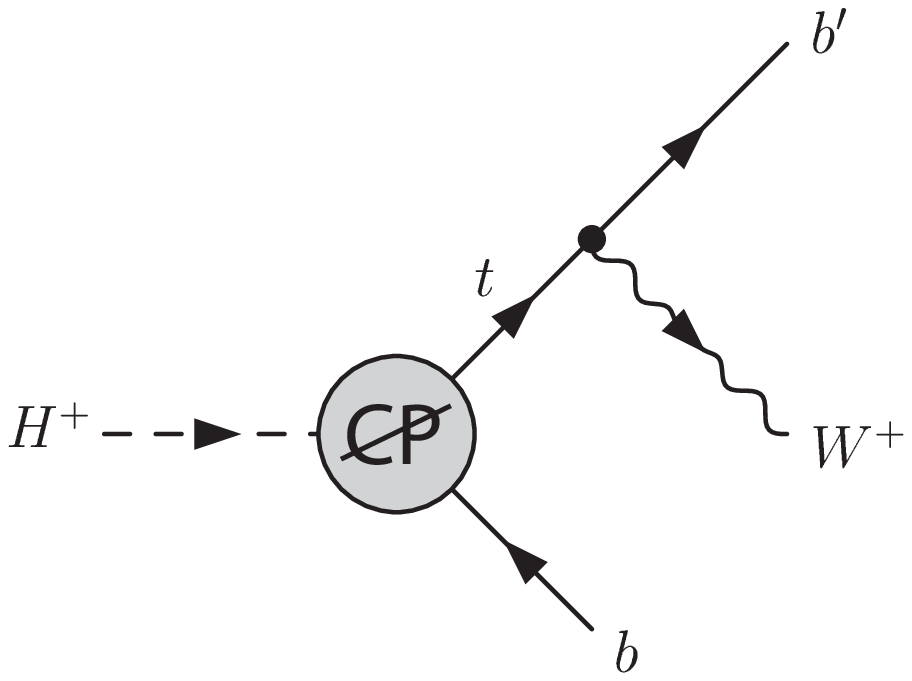}}} \hfil
 \mbox{\resizebox{!}{4.3cm}{\includegraphics{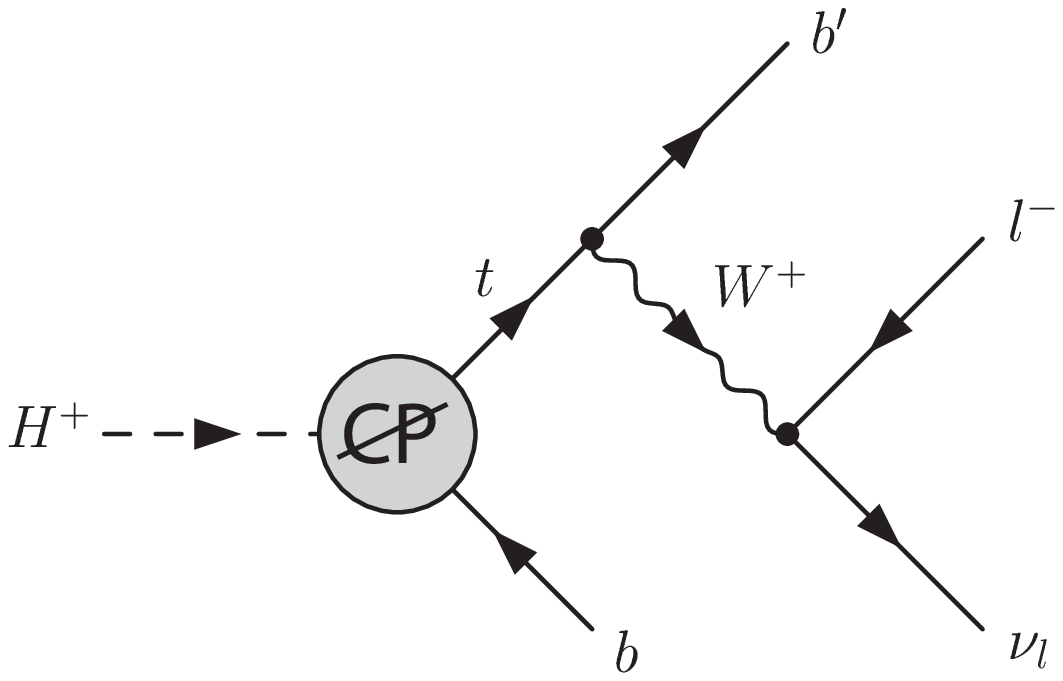}}} 
  \end{center}
  \caption{The Feynman graphs of the processes we study.
  \label{fig:feyn}}
\end{figure}

\begin{figure}[ht]
{\setlength{\unitlength}{1mm}
\begin{center}
\begin{picture}(170,40)
\put(10,0){\mbox{\resizebox{!}{35mm}{\includegraphics{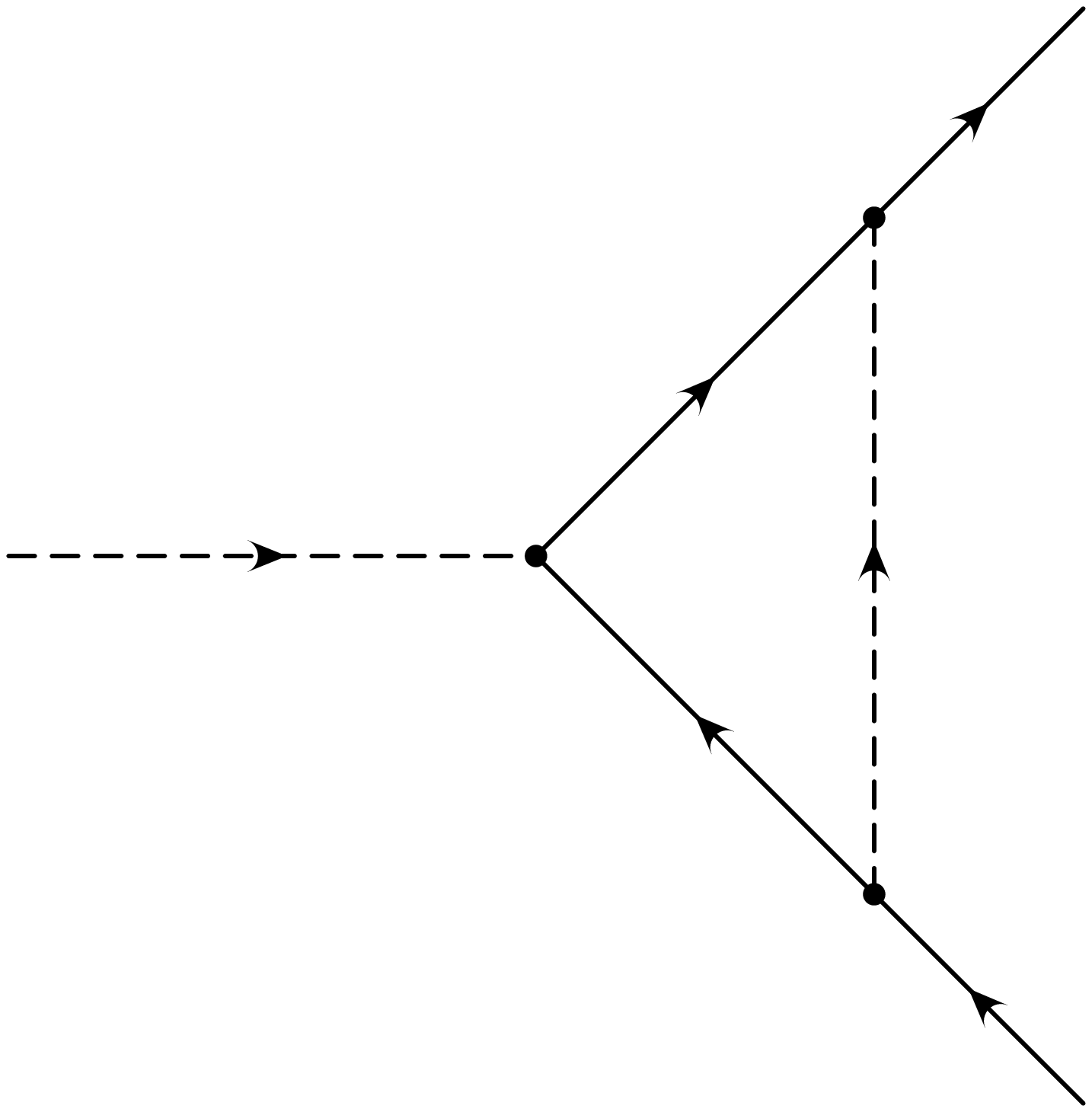}}}}
\put(75,0){\mbox{\resizebox{!}{35mm}{\includegraphics{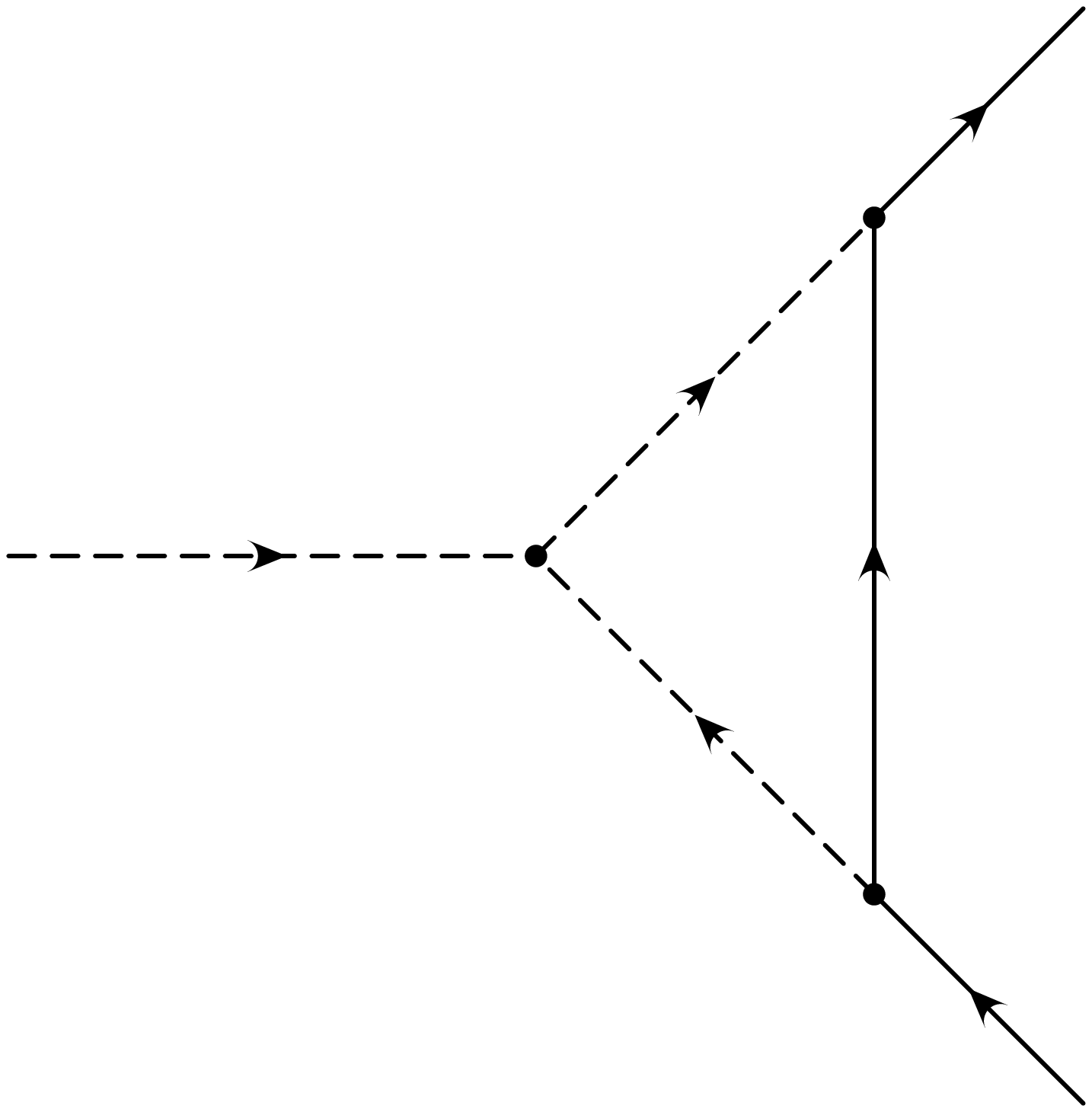}}}}
\put(1,16.5){\mbox{$H^+$}} \put(66,16.5){\mbox{$H^+$}}
\put(47,34){\mbox{$t$}} \put(47,-1){\mbox{$b$}}
\put(112,34){\mbox{$t$}} \put(112,-1){\mbox{$b$}}
\put(16,25){\mbox{$\nt_k\;(\chp_j)$}}
\put(16,8){\mbox{$\chm_j\;(\nt_k)$}}
\put(41,16){\mbox{$\st_i^{}\;(\sb_i^{})$}}
\put(91,25){\mbox{$\st_i$}} \put(91,8){\mbox{$\sb_j$}}
\put(106,16){\mbox{$\nt_k,\,\sg$}}
\put(2,34){\mbox{\bf a)}} \put(67,34){\mbox{\bf b)}}
\end{picture}

\begin{picture}(170,30)
\put(10,0){\mbox{\resizebox{!}{20mm}{\includegraphics{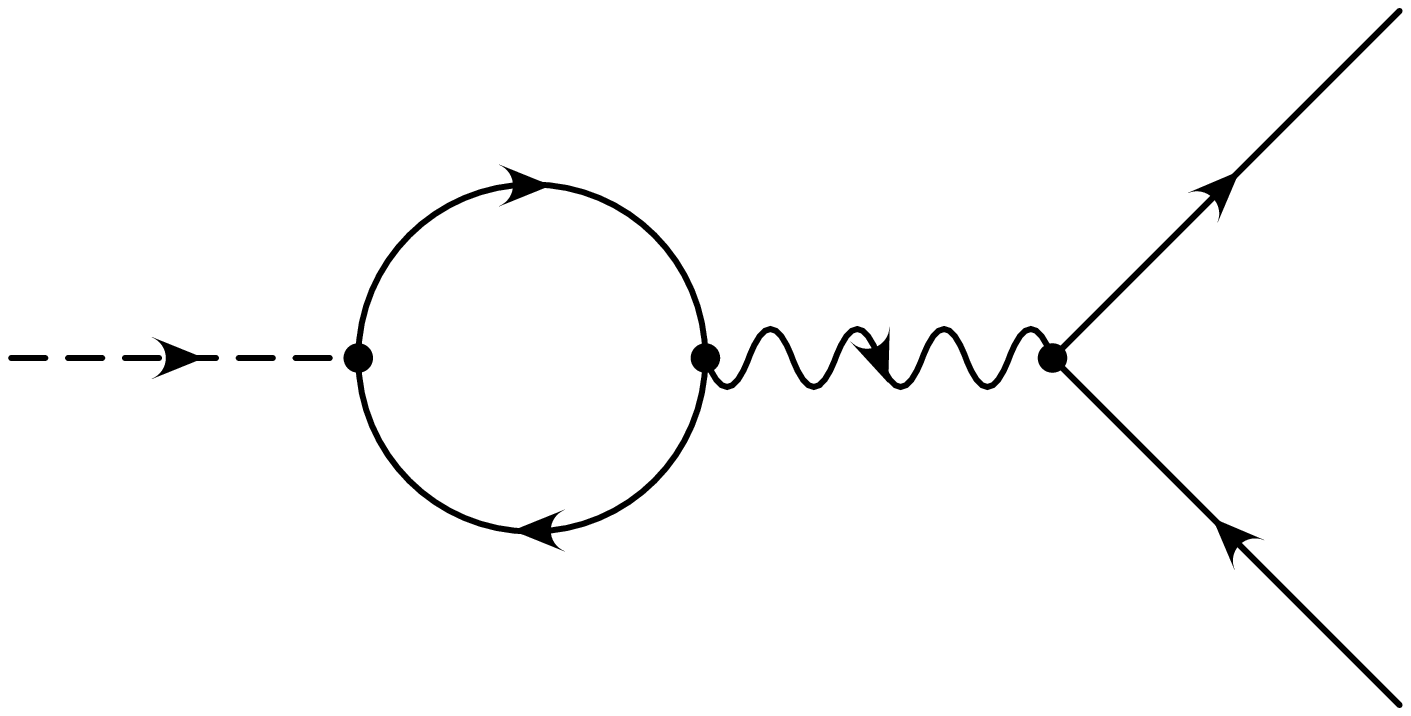}}}}
\put(75,0){\mbox{\resizebox{!}{20mm}{\includegraphics{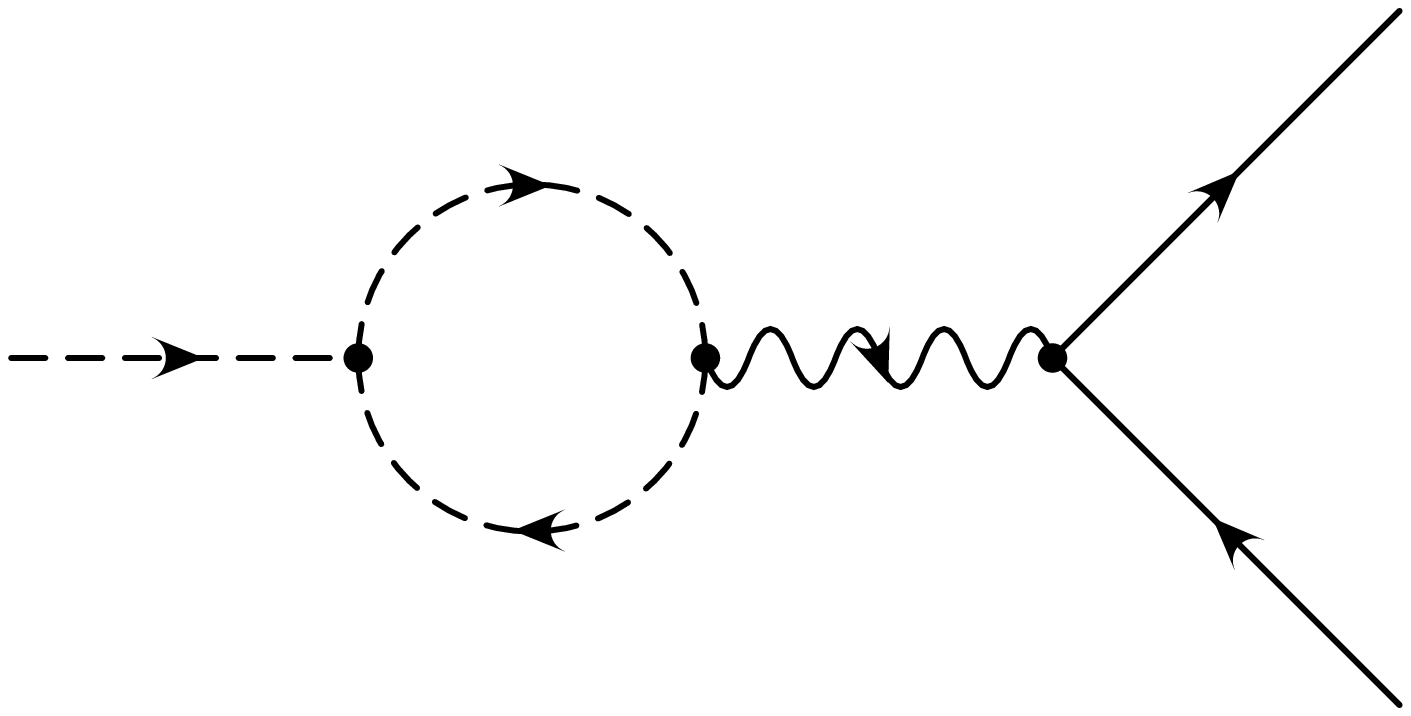}}}}
\put(1,9.5){\mbox{$H^+$}} \put(66,9.5){\mbox{$H^+$}}
  \put(52,19){\mbox{$t$}}
\put(117,19){\mbox{$t$}} \put(52,-1){\mbox{$b$}}
\put(117,-1){\mbox{$b$}}
\put(23.5,18){\mbox{$\nt_k$}} \put(23,0){\mbox{$\chm_j$}}
\put(33,12.5){\mbox{$W^+$}}
\put(85,17.5){\mbox{$\st_i \;(\stau_i)$}}
\put(85,-0.5){\mbox{$\sb_j \;(\snu)$}} \put(98,12.5){\mbox{$W^+$}}
\put(2,22){\mbox{\bf c)}} \put(67,22){\mbox{\bf d)}}
\end{picture}
\end{center}}
\caption{Sources for CP violation in $H^+\to t\bar b$ decays at
1-loop level in the MSSM with complex couplings ($i,j=1,2;$ $k=1,...,4$).
\label{fig:feyngraphs}}
\end{figure}

\begin{figure}[h!]
  \resizebox{8cm}{!}{\includegraphics{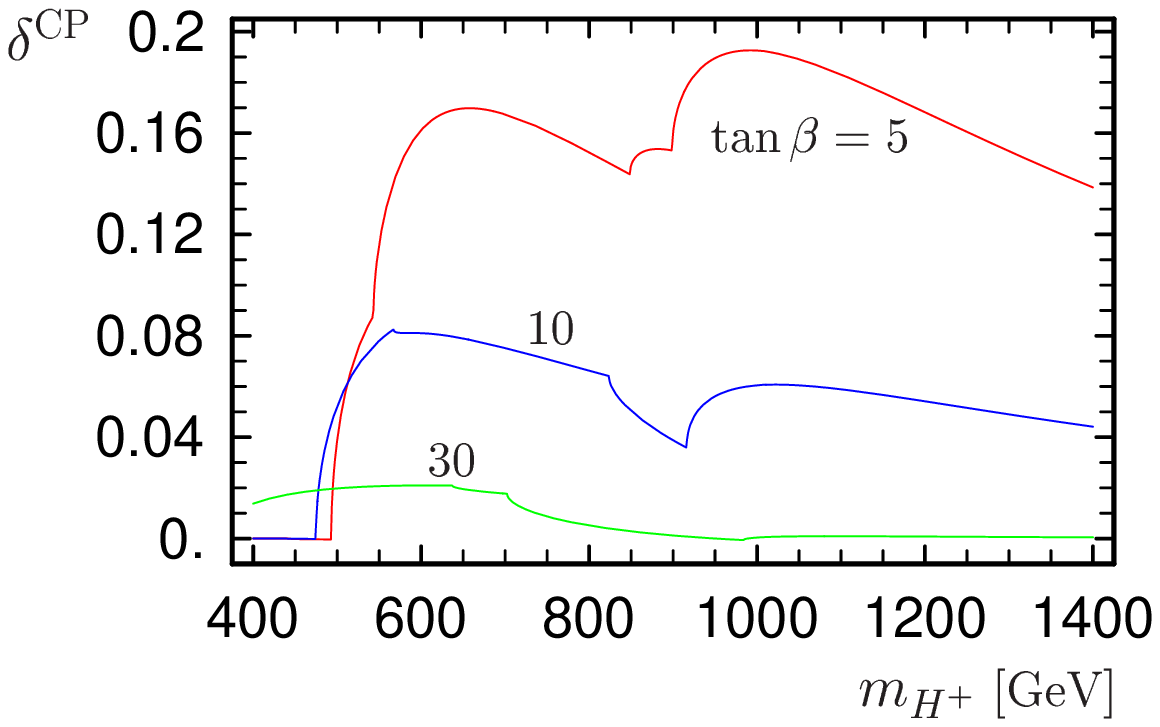}}
  \resizebox{8cm}{!}{\includegraphics{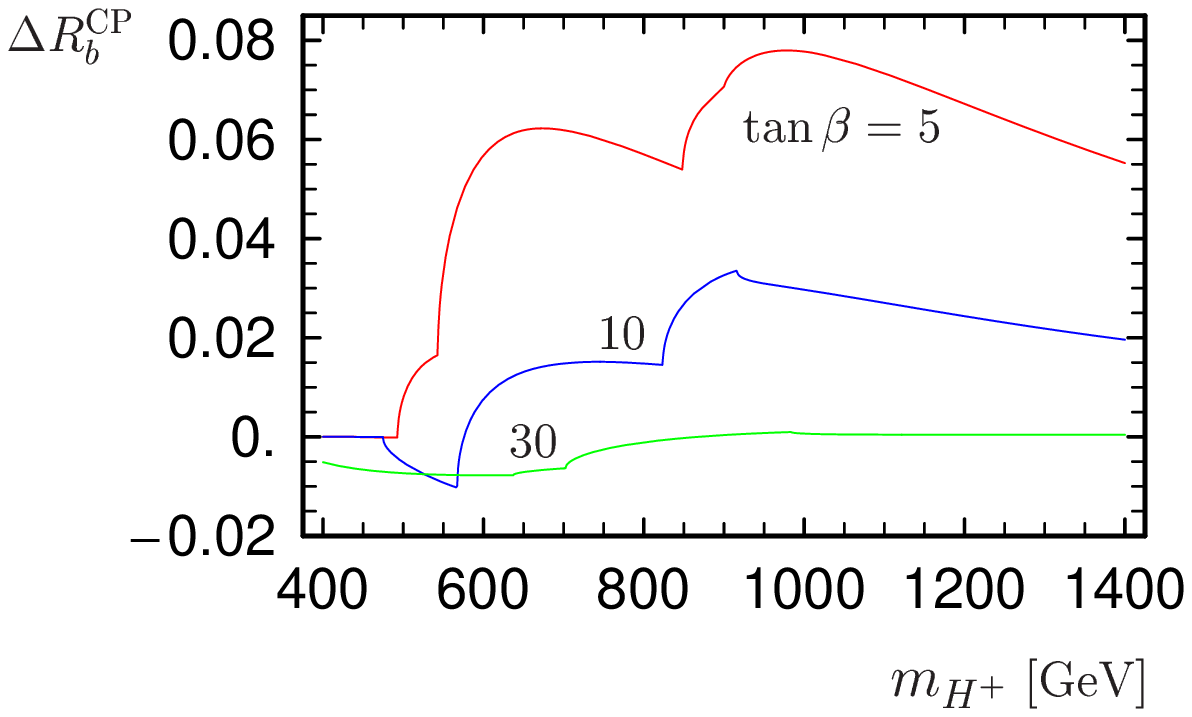}} \\
  \caption{The asymmetries $\delta^{CP}$ and $\Delta R_b^{CP}$ as
 functions of $m_{H^+}$ for  $\phi_{A_t}=\pi/2$,\,
$\phi_{A_b}=\phi_\mu =0$. The red, blue, and green lines are for
$\tan\beta=5,10$, and $30$. \label{fig:2b}}
\end{figure}

\begin{figure}[h!]
  \resizebox{8cm}{!}{\includegraphics{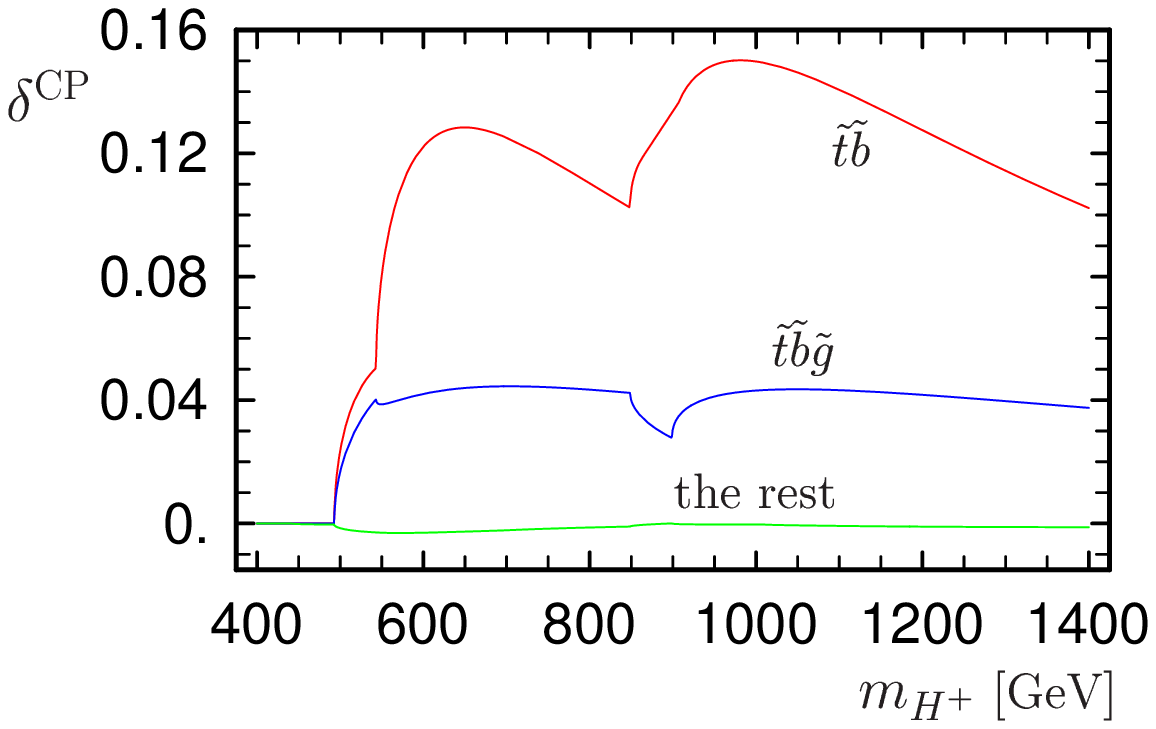}}
  \resizebox{8cm}{!}{\includegraphics{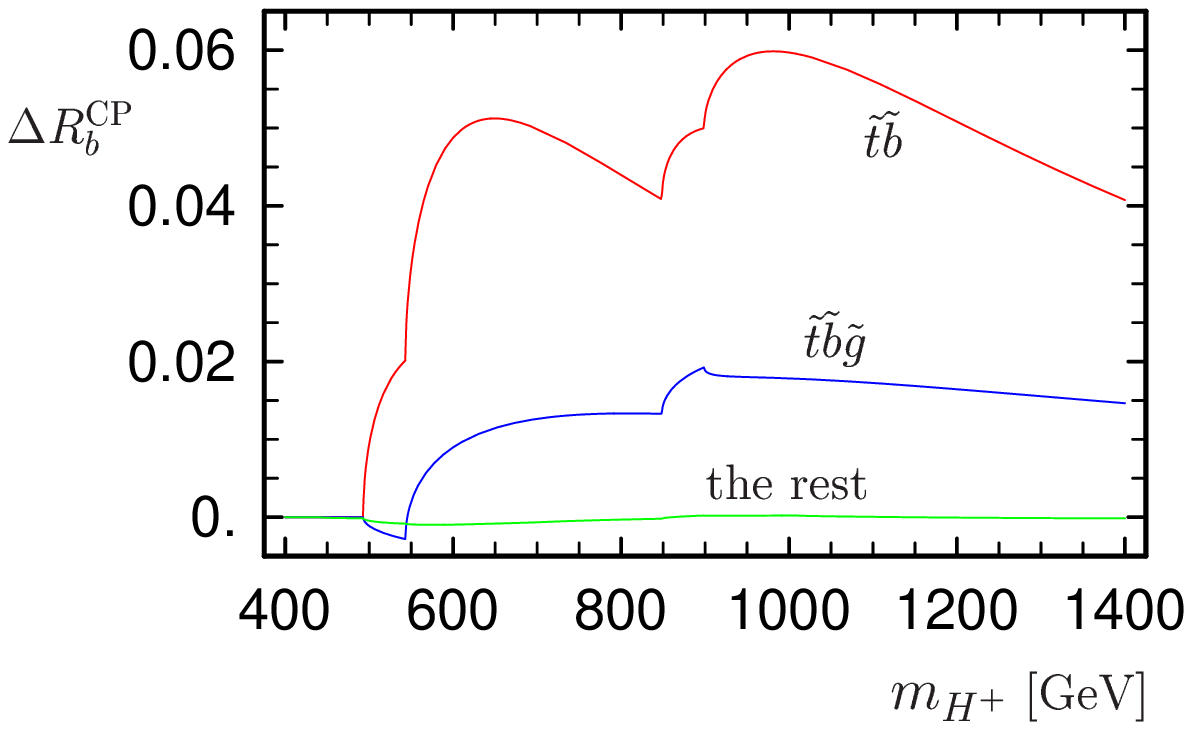}} \\
  \caption{The contribution of the $\tilde t\tilde b$ self-energy
 (red line), $\tilde t\tilde b\tilde g$ vertex contribution (blue line) and
the sum of the other (green line) diagrams to
$\delta^{CP}$ and $\Delta R_b^{CP}$ as functions of $m_{H^+}$ for $\tan\beta =5$
and $\phi_{A_t}=\pi/2$, $\phi_{A_b}=\phi_\mu =0$. \label{fig:3b}}
\end{figure}

\begin{figure}[h!]
  \resizebox{8cm}{!}{\includegraphics{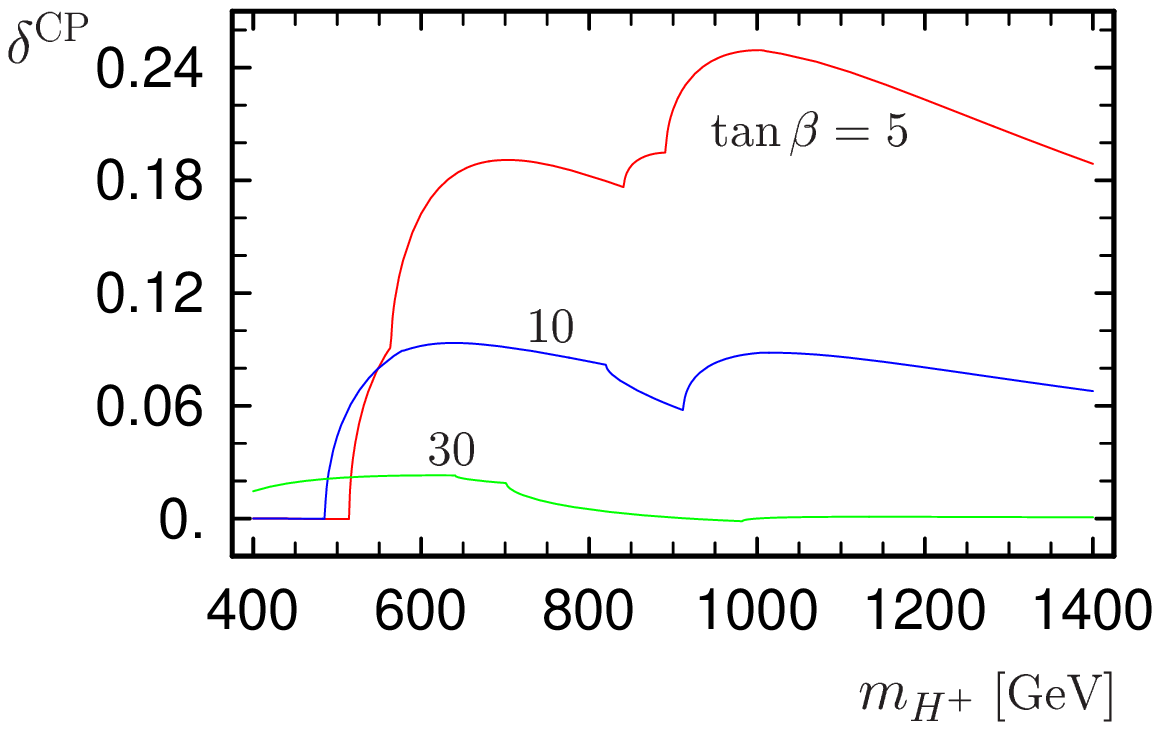}}
  \resizebox{8cm}{!}{\includegraphics{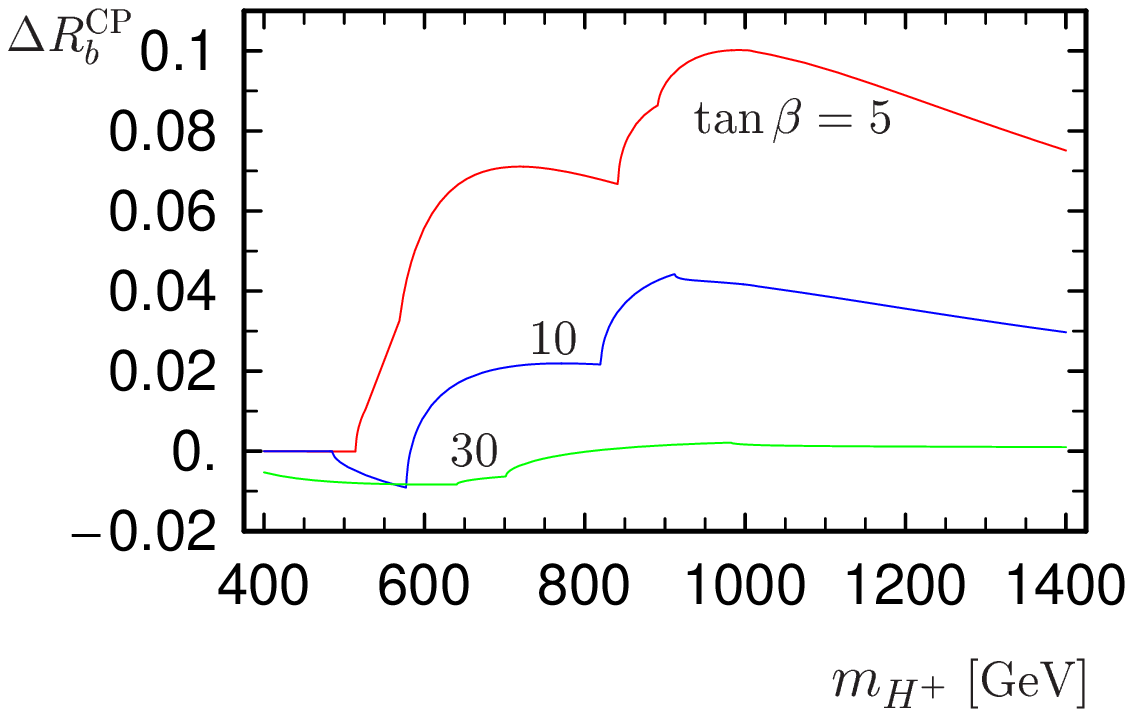}} \\
  \caption{The asymmetries $\delta^{CP}$ and $\Delta R_b^{CP}$ as
 functions of $m_{H^+}$ for  $\phi_{A_t}=\pi/2$, $\phi_{A_b}=0$
and a non zero phase of $\mu$, $\phi_\mu=\pi/10$. The red, blue
and green lines are for $\tan\beta=5,10$, and $30$. \label{fig:5a}}
\end{figure}

We study the following processes, related to the quark decay mode of  $H^\pm$~\cite{second} (See Fig.~\ref{fig:feyn})
\begin{eqnarray}
H^{+}\rightarrow \bar{b}\,t \rightarrow \bar{b}\,b'\,W^{+}\, ,\nn
\\ H^{-}\rightarrow b\,\bar{t} \rightarrow b\,\bar{b}'\,W^{-}\,
,\label{pro1}
\end{eqnarray}
and
\begin{eqnarray}
 H^{+}\rightarrow \bar{b}\,t \rightarrow \bar{b}\,b'\,W^{+} \rightarrow
\bar{b}\,b'\,l^{+}\,\nu_l \, ,\nn \\  H^{-}\rightarrow
b\,\bar{t}\rightarrow b\,\bar{b}'\,W^{-} \rightarrow
b\,\bar{b}'\,l^{-}\,\bar\nu_l\, \label{pro2}.
\end{eqnarray}
The CP--violation is induced by SUSY loop corrections in the $H^\pm tb$--vertices, whose effective amplitudes are given by
\begin{equation}
\label{amp1} {\cal M}_{H^{+}}=i
\bar{u}(p_t)[Y_b^{+}P_R+Y_t^+P_L]u(-p_{\bar b})
\end{equation}
\begin{equation}
\label{amp2} {\cal M}_{H^{-}}=i
\bar{u}(p_b)[Y_t^{-}P_R+Y_b^-P_L]u(-p_{\bar t})\,,
\end{equation}
The one loop diagrams that contribute are shown on \fig{feyngraphs}. The Yukawa couplings 
$ Y_i^\pm$ ($i=t,b$) are a
sum of the tree-level couplings $y_t$ and $y_b$, and the
contributions from the loops -- $\d Y_i^\pm$ ($i=t,b$):
\begin{equation}
  Y_i^{\pm} = y_i^{} + \d Y_i^{\pm} \quad i=t,b \,.\label{sravni1}
\end{equation}
The loop contributions $\d Y_i^\pm$ have, in general, both
CP-invariant and CP-violating parts:
\begin{equation}
\d Y_i^\pm = \d Y_i^{inv} \pm \smaf{1}{2}\,\d Y_i^{CP}\,.
\label{eq:dYi}
\end{equation}
In addition, both CP-invariant and CP-violating parts have real
and imaginary (absorptive) parts. Only the real parts ${\rm Re}\,\d
Y_{t,b}^{CP}$ enter the CP-violating asymmetries.
We examine the following CP-violating asymmetries related
to the processes (\ref{pro1}) and (\ref{pro2}):
\begin{itemize}
\item{{\bf CP-violating decay rate asymmetry  \boldmath
$\delta ^{CP}$}, defined as}
\end{itemize}
\begin{equation}
 \delta ^{CP}_{f}={N_f -N_{\bar f} \over N_f +N_{\bar f}} \, , \qquad f=b',l^\pm
\end{equation}
where $N_{f,\bar f}$ is the total number of particles $f (\bar f)$ in (\ref{pro1}) or  (\ref{pro2}) respectively.
\begin{itemize}
\item{{\bf CP-violating forward-backward (FB) asymmetry
\boldmath $\Delta A_{b(l)}^{CP}$}, which we construct from the FB
asymmetries $ A^{FB}_{f,\pm}$ of the processes (\ref{pro1}) and
(\ref{pro2}) using the angular distributions of the fermions $f=b', l^\pm$ from the top quark decay in (\ref{pro1}) and (\ref{pro2}) . This asymmetry is
defind as}
\end{itemize}
\begin{equation}
\Delta A_f^{CP}= A^{FB}_{f+} -  A^{FB}_{f-} \, ,
\end{equation}
\noi where $A^{FB}_{\pm}$ are the ordinary FB asymmetries of the
processes
\begin{equation}
 A^{FB}_{f,\pm}= {\Gamma_{f,\pm}^F-\Gamma_{f,\pm}^B \over
 \Gamma_{f,\pm}^F+\Gamma_{f,\pm}^B}\, ,
\end{equation}
and we have
\begin{equation}
 \Gamma_{\pm}^F=\int_0^{\pi\over 2} {d\Gamma^{\pm}\over d\cos
\theta}d\cos \theta\,, \quad \quad \Gamma_{\pm}^B=\int_{\pi\over
2}^{\pi}{d\Gamma^{\pm}\over d\cos \theta}d\cos \theta \, ,
\end{equation}
\noi i.e. $\Gamma_{f,\pm}^F$ are the number of particles $f (\bar f)$
measured in the forward direction of the decaying $t$($\bar t$)
quarks and $\Gamma_{f,\pm}^B$ are the number of $f (\bar f)$ measured in the backward direction of the decaying
$t$($\bar t$) quarks.
\begin{itemize}
\item{{\bf CP-violating energy asymmetry \boldmath $\Delta
R_b^{CP}$} can be defined analogously, using the energy
distributions of the processes (\ref{pro1}) and (\ref{pro2}),
namely}
\end{itemize}
\begin{equation}
\Delta R_b^{CP}=R_{b+} -  R_{b-} \, , \label{delR}
\end{equation}
where $R_{\pm}$ are given by
\begin{equation}
\label{eq:R}
 R_{b,\pm}={\Gamma_{\pm} (x>x_0)-\Gamma_{\pm}(x<x_0)
\over \Gamma_{\pm}(x>x_0)+\Gamma_{\pm}(x<x_0)} \, ,
\end{equation}
$x$ is a dimensionless variable proportional to the energy, $x=E_{b'}/m_{H^+}$ and
$x_0$ is any fixed value in the energy interval.\\
The analythical results for the asymmetries are rather simple:
\begin{equation}
  \delta_{b} ^{CP} =\delta^{CP}_l ={\Gamma^{CP}\over
\Gamma^{inv}}\,,\label{treto}
\end{equation}
\begin{equation}
\label{eq:DeltaAbCP} \Delta
A^{CP}_{b(l)}=2\alpha_{b(l)}m_t^2m_H^2{m_H^2-m_t^2
\over(m_H^2+m_t^2)^2}{\cal P}^{CP}\,,
\end{equation}
\label{delr}
\begin{equation}
 \Delta R^{CP}_b= {1\over 2}\alpha_b(m_H^2-m_t^2){\cal
P}^{CP}\,, \label{eq:DeltaRbCP}
\end{equation}
where $\Gamma^{CP}$ and $ {\cal P}^{CP}$ are two linear combinations of the two formfactors:
\begin{equation}
\Gamma^{CP}=   \left[ y_t {\rm Re}(\delta Y_t^{CP})+  (y_b {\rm
Re}(\delta Y_b^{CP}) \right](p_tp_{\bar{b}})- m_t m_b \left[
y_t{\rm Re} (\delta Y_b^{CP})+ y_b {\rm Re}(\delta Y_t^{CP})
\right]\, , \label{GammaCP}
\end{equation}
\begin{equation}
{\cal P}^{CP}=  {y_t {\rm Re}(\delta Y_t^{CP})-  y_b {\rm
Re}(\delta Y_b^{CP})\over (y_t^2 + y_b^2) (p_{t }p_{\bar{b}})- 2
m_t m_b y_t y_b} \, . \label{PCP}
\end{equation}
Here $\Gamma^{CP}$ is the CP-violating part of the relevant $H^\pm$ partial decay rate and $ {\cal P}^{CP}$ is the CP--violating part of the polarization of the top-quark. The coefficient $\alpha_{b(l)}$ determines the sensitivity of the b-quark (lepton) to the polarization of the top quark:
\begin{equation}
\alpha_b={m_t^2-2m_W^2\over m_t^2+2m_W^2}, \qquad \alpha_l=1\,.
\end{equation}
Further analysis on the results shows that
one does not need to explore numerically all of these asymmetries, as some of them are connected. In this
context we reduce the numerical analysis on the following
argumentation:
\begin{itemize}
\item{ All of the discussed asymmetries are linear combinations of
the CP-violating form factors ${\rm Re}(\delta Y_t^{CP})$ and
${\rm Re}(\delta Y_b^{CP})$, through a dependance on ${\cal
P}^{CP}$ (\ref{PCP}) or $\Gamma^{CP}$ (\ref{GammaCP}). In
order to measure these formfactors, one needs to measure: {\bf 1)}
the decay rate asymmetries $\delta_{b,l}^{CP}$ which are
proportional to $\Gamma^{CP}$ and 2) the angular and /
or energy asymmetries which are proportional to ${\cal P}^{CP}$
for a given process.}
\item{ As there is no CP violation in $t\to bW$, the decay rate
asymmetries in $H^\pm\rightarrow W^\pm bb'$ and $H^\pm\rightarrow
bb'l^\pm \nu$ are equal to the decay rate asymmetry in $H^\pm\to
tb$. We denote it, following \cite{we2},  by $\delta^{CP}$:}
\begin{equation}
\qquad
\delta^{CP}=\delta_{b}^{CP}=\delta_l^{CP}=\frac{\Gamma^{CP}}{\Gamma^{inv}}\,. \qquad \delta^{CP}=\frac{\Gamma^+ -\Gamma^-}{\Gamma^+ +\Gamma^-}\,,
\end{equation}
where $\Gamma^\pm$ is the partial decay rate of the process $H^\pm \to tb$
\item{ The angular and energy asymmetries -- $\Delta A_{b,l}^{CP}$
and $\Delta R_b^{CP}$, are not independent either. The angular
asymmetries for the $b$-quarks and the leptons are related by
\begin{equation} \Delta A_l^{CP} = \frac{\alpha_l}{\alpha_b}\,\,
\Delta A_b^{CP} \approx 2.6\, \Delta A_b^{CP}\,.\label{r1}
\end{equation}
Furthermore, there is
a simple relation between the $b$-quark energy and angular
asymmetries:
\begin{equation} \Delta R_b^{CP} = {(m_{H^+}^2 + m_t^2)^2 \over 4\,
m_{H^+}^2\, m_t^2}\, \Delta A_b^{CP}\, ,\label{r2}
\end{equation}
which implies that for $m_{H^+} > m_t$, $\Delta R_b^{CP}$ is
bigger than $\Delta A_b^{CP}$. Thus, in general, $\Delta R_b^{CP}
$ is the biggest asymmetry of those defined through the polarization of the top quark for $m_{H^+} > 490$~GeV.}
\end{itemize}
In the following, we present numerical results only on the
decay rate asymmetry $\delta^{CP}$ and on the energy asymmetry
$\Delta R_b^{CP}$.\\
In order not to vary
too many parameters we fix a part of the parameter space by the
choice:
\begin{eqnarray}
    M_2=300~{\rm GeV}, \qquad M_3=745~{\rm GeV}, \qquad
     M_{\tilde U}=M_{\tilde Q}=M_{\tilde D}=M_E=M_L =350~ {\rm
     GeV},\nn\\
\mu=-700~{\rm GeV},\qquad  |A_t| = |A_b| = |A_\tau|=700~{\rm
GeV}.\qquad \qquad \qquad \qquad\qquad \qquad
\qquad\label{parameters}
\end{eqnarray}
According to the
experimental limits on the electric and neutron EDM's, we take
$\phi_\mu=0$ or $\phi_\mu=\pi /10$. The remaining CP-violating
phases we vary, are the phases of $A_t$,
$A_b$ and $A_\tau$.  In the numerical code we use running top and bottom Yukawa
couplings, calculated at the scale $Q = m_{H^+}$.
Fig.~\ref{fig:2b}a and Fig.~\ref{fig:2b}b show the asymmetries $\delta^{CP}$ and $\Delta
R_b^{CP}$ as functions of $m_{H^+}$ for $\phi_{A_t}=\pi/2$,
$\phi_{A_b}=0$ and $\phi_\mu=0$. (Our studies have shown that the
most important CP-violating phase is $\phi_{A_t}$. There is only a
very weak dependence on $\phi_{A_b}$ and $\phi_{A_\tau}$ and
therefore we take them zero.)  For $\tan\beta=5$ the
decay rate asymmetry $\delta^{CP}$ goes up to $20\%$, while
$\Delta R_b^{CP}$ reaches $8\%$ for the same values of the
parameters. The asymmetries strongly depend on $\tan\beta$ and
they quickly decrease as $\tan\beta$ increases. For $\mhp < \mst{1}+\msb{1}$, the asymmetries are very small. The
main contributions to both $\delta^{CP}$ and $\Delta R_b^{CP}$
come from the self-energy graph with stop-sbottom. The vertex
graph with stop-sbottom-gluino also gives a relatively large
contribution. The contributions of these two graphs in  $\delta^{CP}$ and $\Delta
R_b^{CP}$  are shown in
Fig.~\ref{fig:3b}a and Fig.~\ref{fig:3b}b. The contribution of the rest of the graphs is
negligible.\\
Further, we take a very small phase, $\phi_\mu = \pi/10$, in order to
fit with the experimental data. As can be seen in
Fig.~\ref{fig:5a}a and Fig.~\ref{fig:5a}b, the asymmetries can increase up to $25\%$ for
$\delta^{\rm CP}$ and $10\%$ for $\Delta R_b^{\rm CP}$, respectively.
The discussed asymmetries $\delta^{CP}$ and $\Delta R_b^{CP }$
shows a very strong dependence on the sign of $\mu$. As noted
above, our analysis is done for $\mu = - 700$ (see
(\ref{parameters})), however if $\mu$ changes sign, $\mu =700$,
all asymmetries become extremely small.
\begin{figure}[t]
\includegraphics{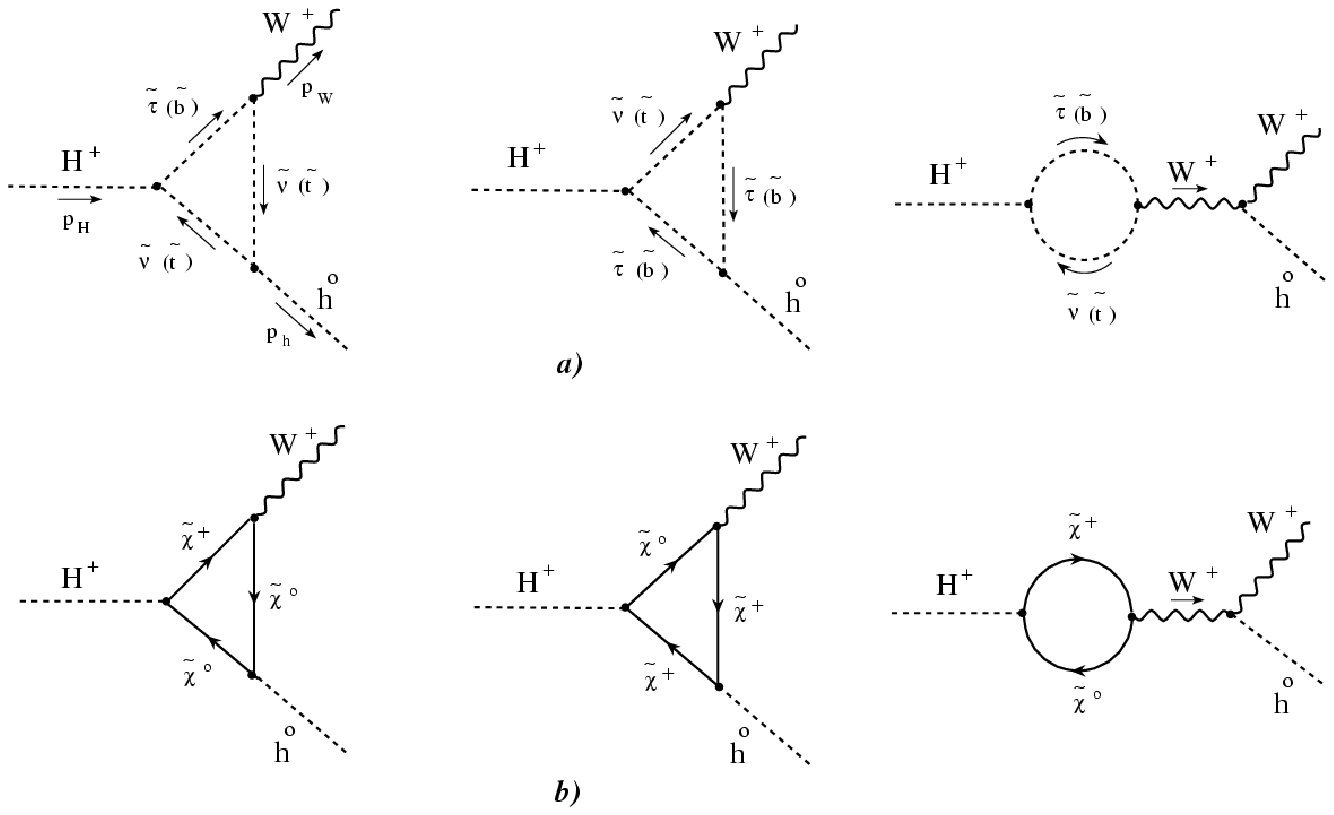} \caption{ The 1-loop diagrams in MSSM
with complex parameters that contribute to
$\delta_{Wh^0}^{CP}$}\label{kartinka}
\end{figure}

We now turn to the bosonic decay mode of the charged Higgs boson
$H^+ \to W^+ h^0$. We are interested again in the CP violation
caused by SUSY loop corrections in the $H^\pm Wh^0$--vertex and study
the decay rate asymmetry between the  charge conjugate processes $H^+
\to W^+ h^0$ and $H^-
\to W^- h^0$:
\begin{equation} \delta_{Wh^0}^{CP} =\frac{\Gamma (H^+ \to W^+h^0) - \Gamma (H^- \to
W^-h^0)} {\Gamma (H^+ \to W^+h^0) + \Gamma (H^- \to
W^-h^0)}.\label{111} \end{equation}
The matrix elements of $H^+\to W^+h^0$ and $H^-\to
W^-h^0$ are given by
 \begin{equation}
M_{H^\pm}=ig\varepsilon_{\alpha}^{\lambda}(p_W)p_h^{\alpha}Y^\pm,
\end{equation}%
where $\varepsilon_{\alpha}^{\lambda}(p_W)$ is the polarization
vector of $W^\pm$, $Y^\pm$ are the loop corrected couplings.
The CP-violating asymmetry $\delta_{Wh}^{CP}$ is determined by the real parts ${\rm Re}
(\delta Y_k^{CP})$:
\begin{equation}
 \delta Y_k^{CP}={\rm Re} (\delta Y_k^{CP})+i\,{\rm Im} (\delta Y_k^{CP})
\,, \qquad  \delta_{Wh^0}^{CP}\simeq
 {2\sum \limits_k {\rm Re}(\delta Y_k^{CP})\over y }\,,\end{equation}
where the sum is over the loops with CP-violation
shown on Fig.~\ref{kartinka}.
We assume that the squarks are heavy and that the decay $H^+ \to
\tilde t \tilde b$ is not allowed kinematically. Thus, we are left with the loop corrections with sleptons, charginos and neutralinos.  In the commonly
discussed models of SUSY breaking, the squarks are much heavier
than  sleptons,  charginos and neutralinos.
According to the
numerical exploration of the $H^\pm \to W^\pm h^0$ decay we have the
following reasoning.\\
 First, increasing $m_{H^+}$, the mass
 $m_{h^0}$ is saturated, approaching its maximum value, $m_{h^0}^{max}\simeq$
 130 GeV. There is an experimental lower bound for $m_{h^0}$, $m_{h^0}\geq 96 $ GeV
 \cite{LEP-h0}.
 Thus, respecting both the experimental and theoretical bounds,
  we consider $m_{h^0}$ in
the range 96 $ \leq m_{h^0}\leq$ 130 GeV. Our analysis shows a very weak
dependence on $m_{h^0}$ and the results here are presented for $ m_{h^0}=125 ~{\rm GeV}$.\\
The second consequence concerns the $H^\pm Wh^0$ coupling which
determines the BR~($H^+\to W^+h^0)$. It
falls down quickly when increasing $m_{H^+}$ and, depending on $\tan\beta$,
 we can enter the so called decoupling limit, $\cos^2 (\beta -\alpha)\to 0$,
 where the BR~($H^+\to W^+h^0)$ almost vanishes. In order to keep the value of BR~($H^+\to W^+h^0)$ at the level of a few percents,
 we keep $m_{H^+}$ and
 $\tan\b$ relatively small: $ 200 \leq m_{H^+}\leq 600 ~{\rm GeV}$ and $3 \leq
\tan\beta \leq 9$.\\
 In
order not to vary too many parameters, we again fix part of the
SUSY parameter space:
 \begin{equation}
  M_2=250~{\rm GeV},~
     M_{ E} = M_{ L} - 5~{\rm GeV},M_L=120~{\rm GeV},
|A_\tau| = 500~{\rm GeV},~
     \vert \mu \vert=150~{\rm GeV.}~\label{leppars}
     \end{equation}
In order $\delta_{Wh^0}^{CP}$ to be nonzero, we need both new
decay channels  opened and CP-violating phases present.
 In accordance with
\begin{figure}[h!]
 \resizebox{8cm}{!}{\includegraphics{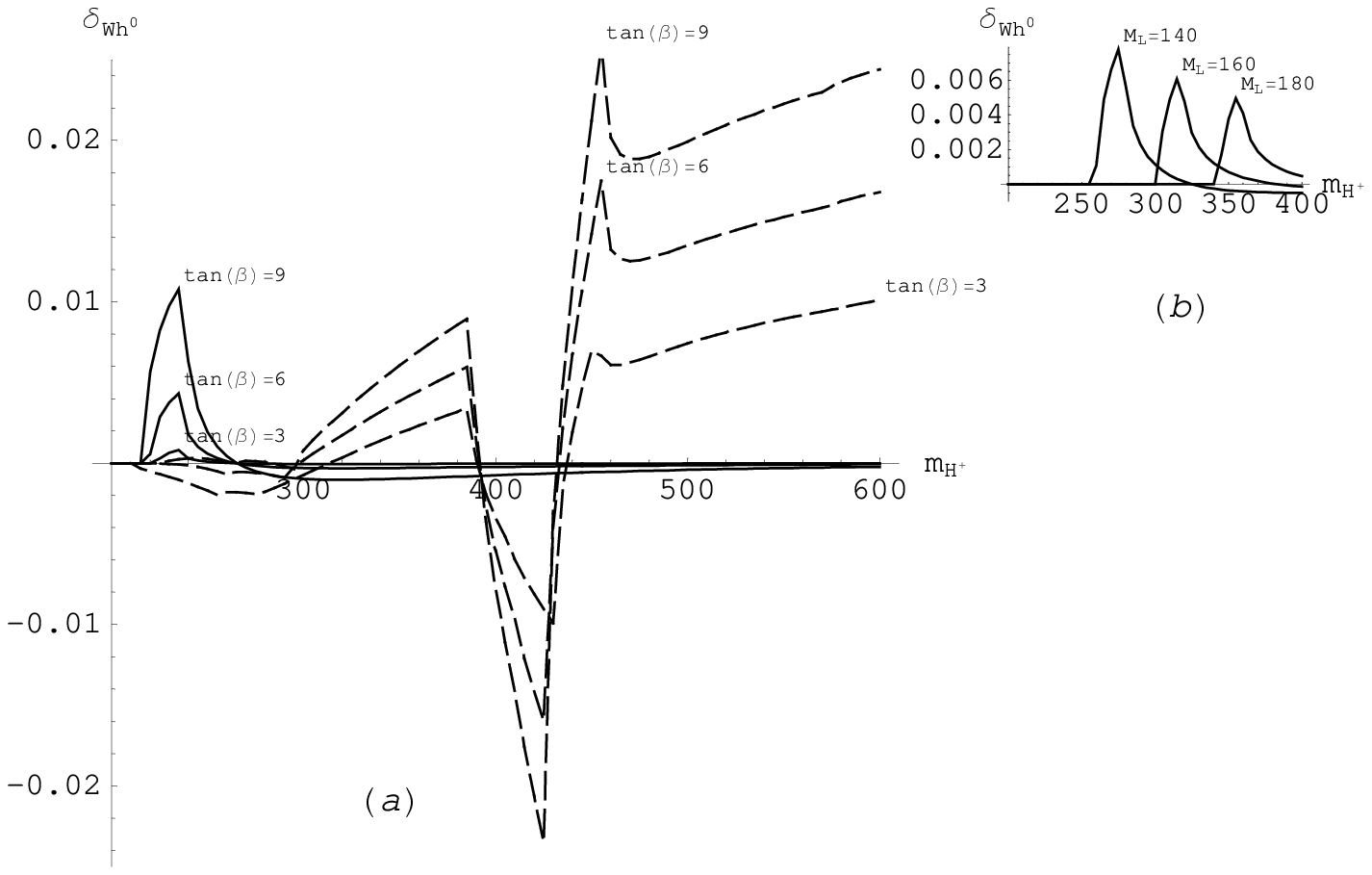}}
 \resizebox{8cm}{!}{\includegraphics{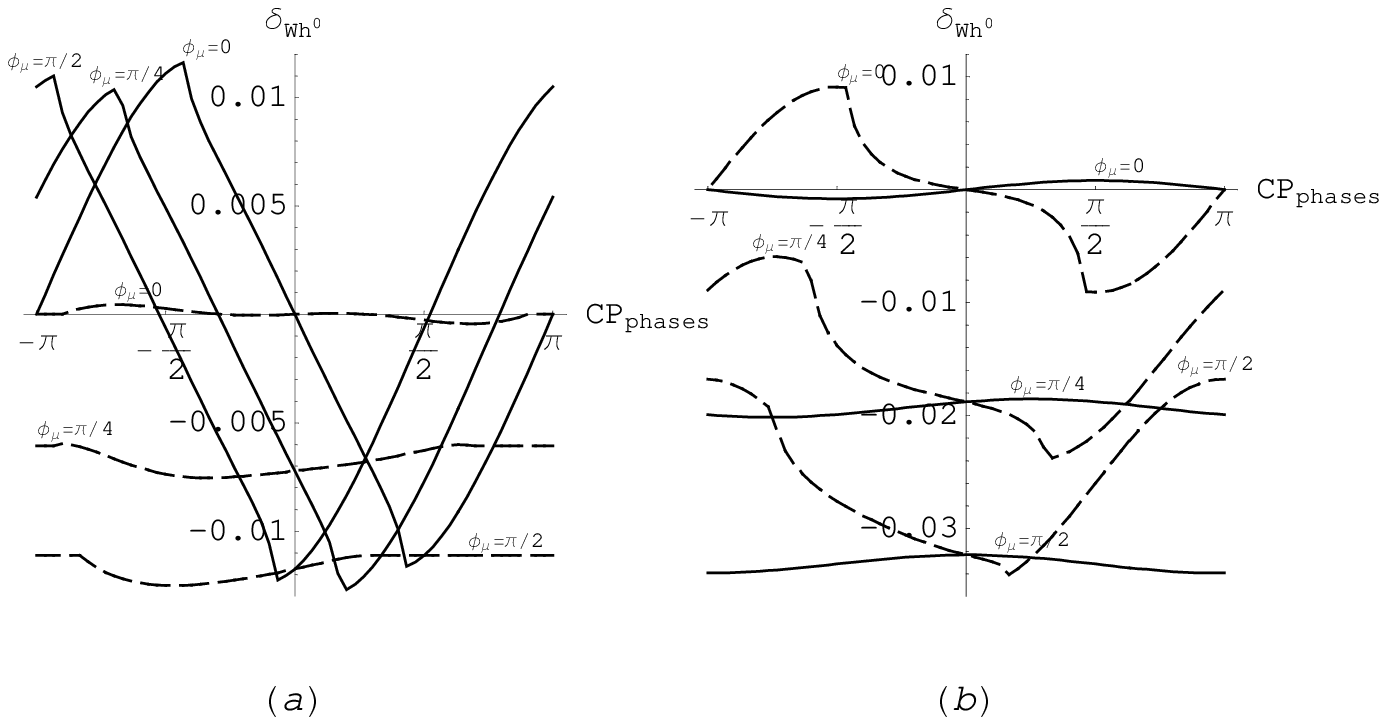}} \\
  \caption{{\bf Left:}$~\delta_{Wh^0}$ as a function of $m_{H^+}$ {\bf a)~} for
different values of $\tan\beta$, $M_L=120$~GeV; solid lines are
for $\phi_\tau=-\pi/2,~\phi_1=0$; dashed lines are for
$\phi_\tau=0,~\phi_1=-\pi/2$. {\bf b)~} for different values of
$M_L$, at $\tan\beta=9,~\phi_\tau=-\pi/2,~\phi_1=0$. {\bf Right:} $~\delta_{Wh^0}$ at $\tan\beta=9$ versus the CP violating
phases $\phi_\tau$ and $\phi_1$ for different values of $\phi_\mu$
[ $\phi_\mu = 0, \pi /4, \pi /2$]. The solid lines are for
$\phi_\tau = [-\pi , \pi ]$ while $\phi_1=0$; the dashed lines are
for $\phi_1 = [-\pi , \pi ]$ while $\phi_\tau =0$,
 {\bf a)~} for $m_{H^+}=237$~GeV
($=m_{\tilde\nu}+m_{\tilde\tau^+_2})$ and
 {\bf b) ~}for
 $m_{H^+}=387$~GeV
($=m_{\tilde\chi^+_2}+m_{\tilde\chi^0_1})$. \label{f1}}
\end{figure}
this we have three cases: {\bf 1)} When only the decay channels
$H^+\to \tilde\nu \tilde\tau^+_n$ are open (Fig.
~\ref{kartinka}a). Then the phase $\phi_\tau$ is responsible for
CP violation. {\bf 2)} When the decay channels
$H^+\to\tilde\chi^+_i \tilde\chi^0_k$ are open only (Fig.
~\ref{kartinka}b). In this case CP violation is due to the phase
$\phi_1$, and {\bf 3)} When both $H^+\to \tilde\nu \tilde\tau^+_n$
and $H^+\to\tilde\chi^+_i \tilde\chi^0_k$ decay channels are
kinematically allowed ( all diagrams of Fig.~\ref{kartinka}). In
this case the two phases $\phi_\tau$ and $\phi_1$ contribute.\\
Examples of the asymmetry as function of $m_H^+$ for cases {\bf
1)} and {\bf 2)} are shown on Fig.~\ref{f1}a for  different values
of $\tan\beta$.  It is clearly seen that in both cases
 the asymmetry strongly increases with $\tan\beta$. In both cases, {\bf 1)} and {\bf
2)}, the asymmetry reaches up to $10^{-2}$. The dependence on
$M_L$ for case {\bf 1)} is seen on Fig.~\ref{f1}b.  Case {\bf 3)},
when all relevant SUSY particles can be light, is described  by
 the algebraic sum of the two graphs at a given $\tan\beta$
 and we don't present it separately. In all these cases the asymmetry does not exceed
 a few percents.\\

 The effect of a non-zero phase $\phi_\mu$ is seen on Fig.~\ref{f1}. In all cases a CP violating phase of $\mu$
 does not change the form of the curves but rather  shifts
 the positions of the maxima and, in general, increases the absolute value of the asymmetry.

Let us summarize. We have calculated different asymmetries concerning the quark and boson decay mode of the charged Higgs boson $H^\pm$, caused by
CP-violating phases in the MSSM Lagrangian. The quark decay mode is dominant for large $m_{H^+}$ and the most important phase is the phase of $A_t$. In this case the decay rate asymmetry can reach up to $25\%$, and the forward-backward asymmetry can go up to $10\%$. These asymmetries are rather large and in principle  measureable at LHC, but because of the large background of the process, require higher luminosity (e.g. SLHC). The boson decay mode is important for small $m_{H^+}$ and $\tan \beta$ and sensitive to the phases of $M_1$ and the phase of $A_{\tau}$. The decay rate asymmetry is, typically, of the order of $10^{-2}\div 10^{-3}$. Concerning its measurability more promising are the next generation linear $e^+e^-$ colliders.


\begin{center}
{\bf \Large Acknowledgements}
\end{center}
This work would not be possible without the valuable contribution of my coauthors Prof. Ekaterina Christova, Prof. Walter Majerotto,
Dr. Helmut Eberl and Dr. Mihail Stoilov.  I am gratefull to all of them.


\end{document}